\documentstyle[11pt,ysc,twoside,epsf]{article}
\def\edcomment#1{\iffalse\marginpar{\raggedright\sl#1\/}\else\relax\fi}
\reversemarginpar

\begin{document}
\title{Accretion to a Magnetized Neutron Star in the "Propeller" Regime}

\author{O.D. Toropina}
\affil{Space Research Institute, Profsojuznaja str. 84/32, Moscow,
Russia,117899}

\author{M.M. Romanova}
\affil{Department of Astronomy, Cornell University, USA}

\author{R.V.E. Lovelace}
\affil{Department of Astronomy, Cornell University, USA}

\begin{abstract}
We investigate spherical accretion to a rotating magnetized star
in the "propeller" regime using axisymmetric resistive
magnetohydrodynamic simulations. The regime is predicted to occur
if the magnetospheric radius is larger than the corotation radius
and smaller  than the light cylinder radius. The simulations show
that accreting matter is expelled from the equatorial region of
the magnetosphere and that it moves away from the star in a
supersonic, disk-shaped outflow. At larger radial distances the
outflow slows down and becomes subsonic. The equatorial matter
outflow is initially driven by the centrifugal force, but at
larger distances the pressure gradient force becomes significant.
We find the fraction of the Bondi accretion rate which accretes to
the surface of the star.

\end{abstract}

\section{Introduction}

Rotating magnetized neutron stars pass through different stages in
their evolution (Shapiro \& Teukolsky 1983, Lipunov 1992).
Initially, a rapidly rotating ($P \le 1{\rm s}$) magnetized neutron
star is expected to be active as a radio-pulsar. The star spins down
owing to the wind of magnetic field and relativistic particles from
the region of the light cylinder $r_L$ (Goldreich \& Julian 1969).
However, after the neutron star spins-down sufficiently, the light
cylinder radius becomes larger than magnetospheric radius $r_m$
where the ram pressure of external matter equals the magnetic
pressure in the neutron star's dipole field. The relativistic wind
is then suppressed by the inflowing matter (Shvartsman 1970). The
external matter may come from the wind from a binary companion or
from the interstellar medium for an isolated neutron star. The
centrifugal force in the equatorial region at $r_m$ is much larger
than gravitational force if $r_m$ is much larger than the corotation
radius $r_{cor}$. In this case the  incoming matter tends to be
flung away from the neutron star by its rotating magnetic field.
This is the so called ``propeller" stage of evolution (Davidson \&
Ostriker 1973, Illarionov \& Sunyaev 1975).

The ``propeller" stage of evolution, though important, is  still
not well-understood theoretically. We discuss results of
axisymmetric, two-dimensional, resistive MHD simulations of
accretion to a a rotating magnetized star in the ``propeller"
regime. We treat the case when matter accretes spherically with
the Bondi accretion rate (Bondi 1952). Bondi accretion to a
non-rotating and a slowly rotating star was investigated by
Toropin et al. (Toropin at al. 1999) and Toropina et al. (Toropina
at al. 2003). Investigation of the accretion to a rotating star in
the ``propeller" regime was started by Romanova et al. (Romanova
et al. 2003).

\begin{figure}
\plotone{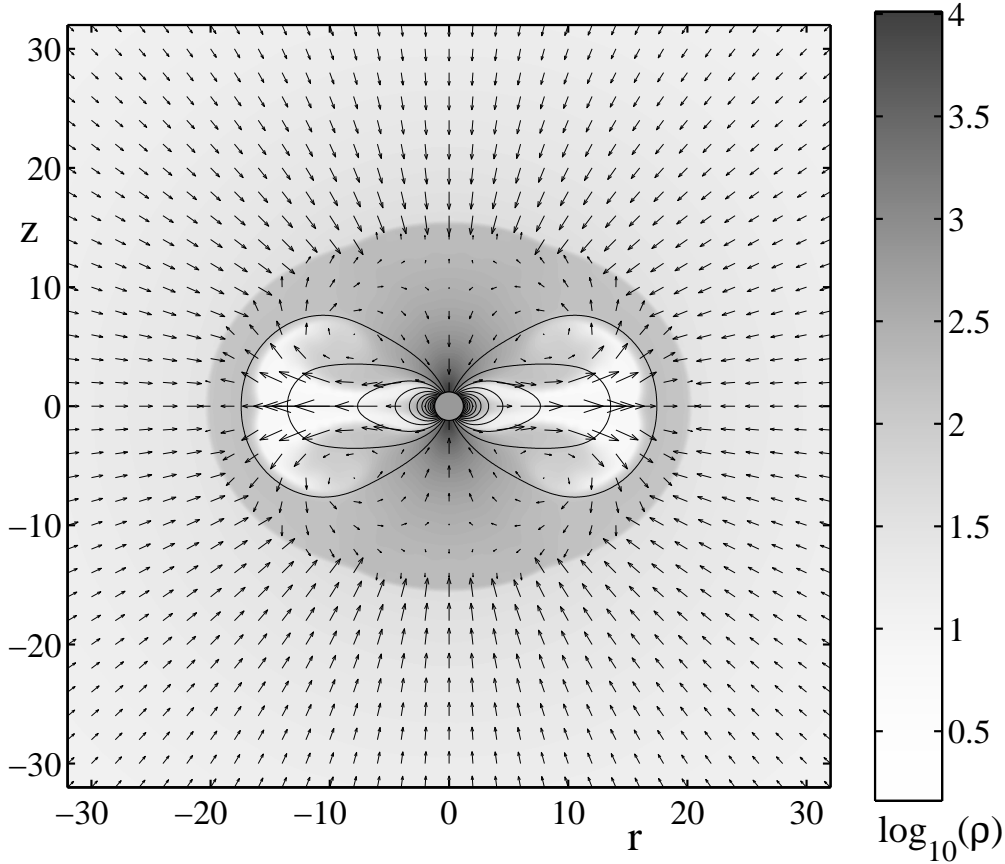} \caption{Matter flow in the ``propeller"
regime for a star rotating at $\Omega_*=0.5\Omega_{K*}$ after $6.9$
rotation periods of the star. The axes are measured in units of the
star's radius. The background represents that the density and the
length of the arrows is proportional to the poloidal velocity. The
thin solid lines are magnetic field lines.}
\end{figure}

\section{Model}

We simulate the plasma flow in the propeller regime using an
axisymmetric, resistive MHD code. The code incorporates the
methods of local iterations and flux-corrected-transport. The code
is described in our earlier  investigation of Bondi accretion to a
non-rotating magnetized star (Toropin at al. 1999, Toropina at al.
2003).
     The equations for resistive MHD are
$$
   {\partial \rho \over
   \partial t}+
   {\bf \nabla}{\bf \cdot}
\left(\rho~{\bf v}\right) =0{~ ,} \eqno(1) $$ $$
\rho\frac{\partial{\bf v}} {\partial t}
             +\rho ({\bf v}
\cdot{\bf \nabla}){\bf v}=
   -{\bf \nabla}p+
{1 \over c}{\bf J}\times {\bf B} + {\bf F}^{g}{ ~,} \eqno(2) $$ $$
   \frac{\partial {\bf B}}
{\partial t} =
   {\bf \nabla}{\bf \times}
\left({\bf v}{\bf \times} {\bf B}\right)
   +
   \frac{c^2}{4\pi\sigma}
\nabla^2{\bf B} {~,} \eqno(3) $$ $$
   \frac{\partial (\rho\varepsilon)
}{\partial t}+
   {\bf \nabla}\cdot \left(\rho
\varepsilon{\bf v}\right) =
   -p\nabla{\bf \cdot}
{\bf v} +\frac{{\bf J}^2} {\sigma}{~.} \eqno(4) $$ We assume
axisymmetry $(\partial/\partial \phi =0)$, but calculate all three
components of ${\bf v}$ and ${\bf B}$.
   The equation of state is taken to be that for an ideal gas,
$p=(\gamma-1)\rho \varepsilon$, with specific heat ratio
$\gamma=7/5$.
     The equations incorporate Ohm's law ${\bf
J}=\sigma({\bf E}+{\bf v}  \times {\bf B}/c)$, where $\sigma$ is
the electrical conductivity.
    The associated magnetic diffusivity,
$\eta_m \equiv c^2\!/(4\pi\sigma)$, is considered to be a constant
within the computational region. In equation (2) the gravitational
force, ${\bf F}^{g} = -GM\!\rho{\bf R}/\!R^3$, is due to the
central star.

We use a cylindrical, inertial coordinate system
$\left(r,\phi,z\right)$ with the $z-$ axis parallel to the star's
dipole moment ${\bf \mu}$ and rotation axis ${\bf \Omega}$. The
equatorial plane is treated as symmetry plane. The vector
potential $\bf A$ is calculated so that ${\bf \nabla}\cdot{\bf
B}=0$ at all times. The star rotates with angular velocity ${\bf
\Omega}_*=\Omega_* ~ \hat{\bf z}$. The intrinsic magnetic field of
the star is taken to be an aligned dipole, ${\bf B} =[3{\bf
R}\left({\bf \mu}\cdot {\bf R} \right)-R^2 {\bf \mu} ]/{R^5}$ with
${\bf \mu} =\mu ~\hat{\bf z}$ and vector-potential ${\bf A}={\bf
\mu} \times{\bf R}/{R^3}$.

We measure length in units of the Bondi radius $R_B \equiv
{GM}/c_\infty^2$, with $c_\infty$ the sound speed at infinity,
density in units of the density at infinity $\rho_\infty$, and
magnetic field strength in units of $B_0$ which is the field at
the pole of the numerical star ($r=0, z=Z_*$). Pressure is
measured in units of $B_0^2/8\pi$. The magnetic moment is measured
in units of $\mu_0 =B_0 R_B^3/2$.

Simulations were done in a cylindrical region $(0\le z\le
Z_{max},~ 0\le r\le R_{max})$. The size of the region is less then
the sonic radius of the Bondi flow $R_{s}$:
$R_{max}=Z_{max}=R_s/\sqrt{2}$. Thus matter inflows supersonically
to the computational region. The inflow rate is taken to be the
Bondi accretion rate, $\dot{M}_B=4\pi\lambda(GM_{\star})^2
\rho_{\infty}/c_\infty^3$, where $\lambda=0.625$ for $\gamma=7/5$.
A uniform $(r,z)$ grid with $513 \times 513$ cells was used.

Initially, the density $\rho(r,z)$  and the velocity ${\bf
v}(r,z)$ are taken to be the values given by the Bondi solution
(Bondi 1952) . Thus, initially $v_\phi =0$. Also, initially the
vector potential ${\bf A}$ was taken to that of a dipole so that
$B_\phi=0$. The star was initialized to be rotating at the rate
$\Omega_*$.

At the outer boundaries, $r=R_{max}$ and $z=Z_{max}$, the
variables $(\rho,~v_r,~v_z,~\varepsilon)$ are fixed and equal to
the values for the Bondi flow which has $v_\phi=0$. The inflowing
matter is unmagnetized with ${\bf B}=0$. At the inner boundary,
the vector-potential ${\bf A}=\hat{\bf \phi~}A_\phi$ of the
intrinsic dipole field of the star is  determined on the surface
of the model star. More detailed description of numerical model
you can see in Romanova et al. 2003.

\section{Results}

\begin{figure}
\plotone{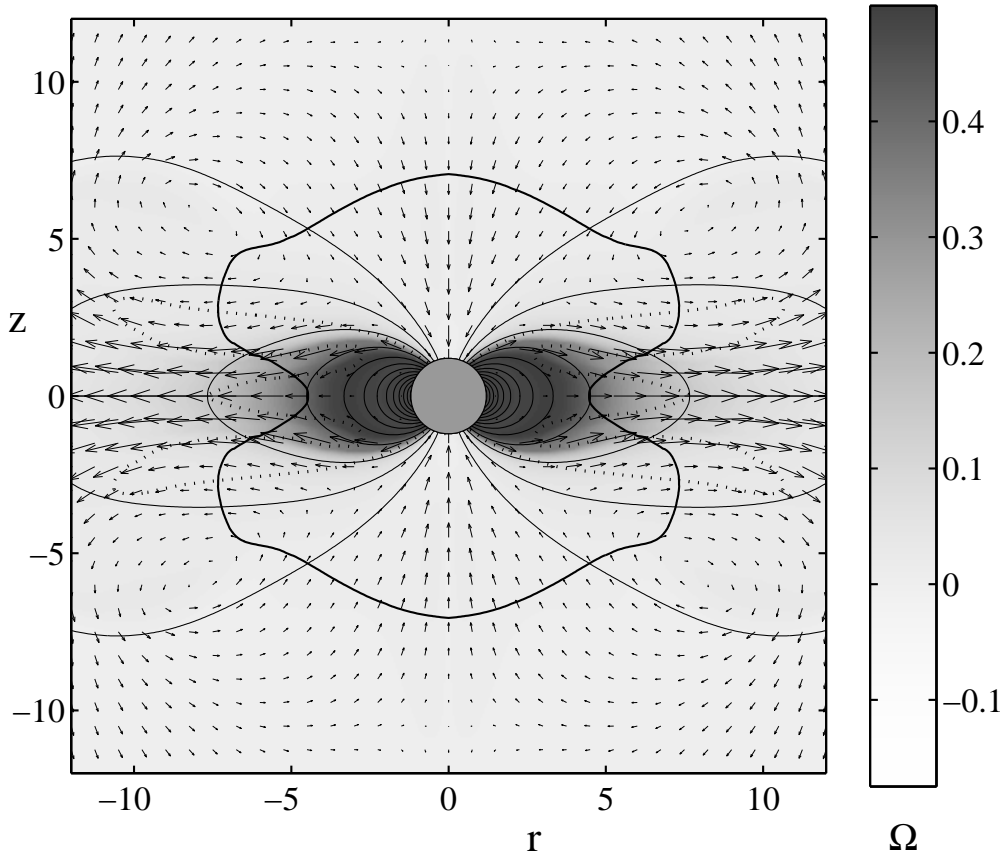} \caption{Enlarged view of Figure 1, the
background represents angular velocity $\Omega=v_\phi(r,z)/r$. The
bold line represents the Alfv\'en surface. Dotted line shows sonic
surface.}
\end{figure}

Here, we first discuss simulations for the case $\omega_*=
\Omega_*/\Omega_{K*}=0.5$, $\beta=10^{-7}$, magnetic diffusivity
$\tilde{\eta}_m=10^{-5}$. For the mentioned values of $\beta$ and
$\eta_m$, the magnetospheric radius $r_{m0}$ and corotation radius
$r_{cor}=(GM/\Omega_*^2)^{1/3}$ are equal for $\omega_* \approx
0.16$. For smaller angular velocities, $\omega_* < 0.16$, the matter
flow around the star is close to that in the non-rotating case. For
$\omega_* > 0.16$ the flow exhibits the features expected in the
``propeller" regime.

Figures 1 shows the general nature of the flow in the propeller
regime. Two distinct regions separated by a shock wave are
observed: One is the {\it external} region where  matter inflows
with the Bondi rate and the density and velocity agree well with
the Bondi solution.
    The second is the {\it internal} region, where the flow is
strongly influenced by the stellar magnetic field and rotation.
The shock wave, which divides these regions, propagates outward as
in the non-rotating case Toropin at al. 1999, Toropina at al.
2003).
   For a rotating star in the propeller regime the shock wave has the
shape of an ellipsoid flattened along the rotation axis of the
star. The region of the flow well within the shock wave is
approximately time-independent. The accretion rate to the star
becomes constant after about $1-2$ rotation periods of the star.

\begin{figure}
\plotone{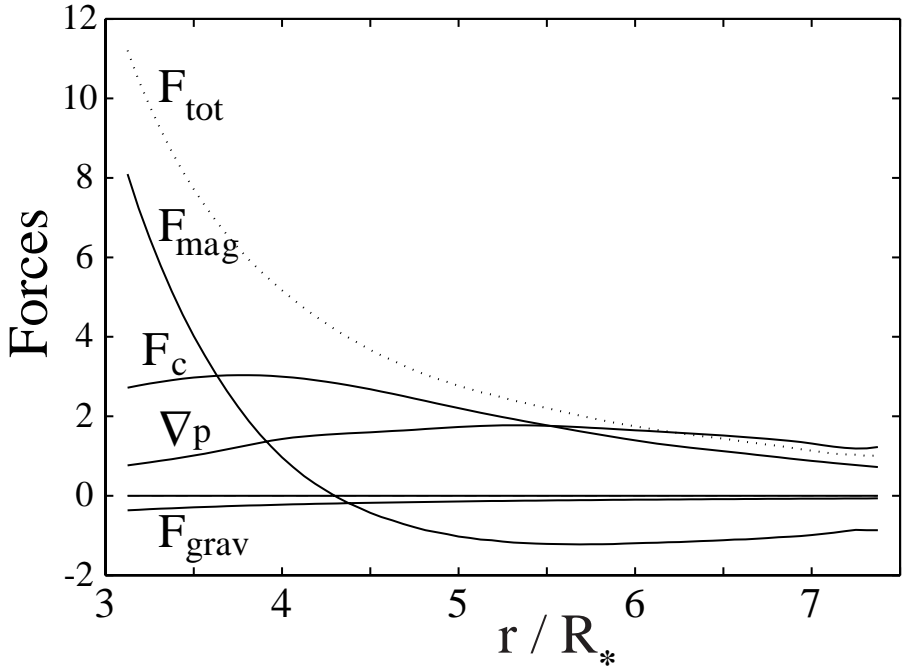} \caption{The radial dependence of the
different radial forces acting in the equatorial plane; ${\rm
F}_{\rm mag}=({\bf J \times B})_r/c$ is the magnetic force (per
unit volume), ${\rm F_c}=\rho v_\phi^2/r$ is the centrifugal
force, ${\bf \nabla}p=(\nabla p)_r$ is minus the pressure gradient
force, ${\rm F}_{\rm grav}=-\rho({\bf \nabla}\Phi_{\rm grav})_r$
is the gravitational force, and ${\rm F}_{\rm tot}$ is the total
radial force.}
\end{figure}

A new regime of matter flow forms inside the expanding shock wave.
The rapidly rotating magnetosphere expels matter outward in the
equatorial region. This matter flows radially outward forming a
low-density rotating torus. The outflowing matter is decelerated
when it reaches the shock wave.

There, the flow changes direction and moves towards the rotation
axis of the star. However, only a small fraction of this matter
accretes to the surface of the star. Most of the matter is expelled
again in the equatorial direction by the rotating field of the star.
Thus, most of the matter circulates inside this inner region driven
by the rapidly rotating magnetosphere.

The solid line in Figure 2 shows the Alfv\'en surface, where the
magnetic energy-density equals the thermal plus kinetic
energy-density, ${\bf B}^2/{8\pi}=\varepsilon + \rho {\bf v}^2/2$.
In the equatorial plane, the Alfv\'en radius is $r_A \approx
4.5R_*$. For $r < r_A$, the magnetic pressure dominates, and the
magnetic field is approximately that of a dipole. At larger
distances the field is stretched by the outflowing plasma. Matter
inside the magnetopause rotates with the angular velocity of the
star.

Figure 3 shows the radial dependencies of the different radial
forces in the equatorial plane. One can see that for $r>3.5 R_*$,
the centrifugal force becomes dominant in accelerating the matter
outward. However, at larger distances, $r > 5.5 R_*$, the pressure
gradient force become larger and determines acceleration of matter.
Thus, centrifugal and pressure gradient forces accelerate matter in
the radial direction. Note, that in most of the region ($r > 4.3
R_*$) the magnetic force is negative so that it opposes the matter
outflow.

The results we have shown are for a relatively strong propeller.
If the rotation rate is smaller, the matter outflow become less
intense. For $\Omega_* < 0.16$, no equatorial outflow is observed
and the shock wave becomes more nearly spherical as it is in the
non-rotating case. On the other hand the shock wave becomes more a
more flattened ellipsoid for larger $\Omega_*$. Matter rotates
rigidly inside the Alfv\'en radius $r_A$ while at large distances
$\Omega= v_\phi/r$ decreases, see Figure 4.

\begin{figure}
\plotone{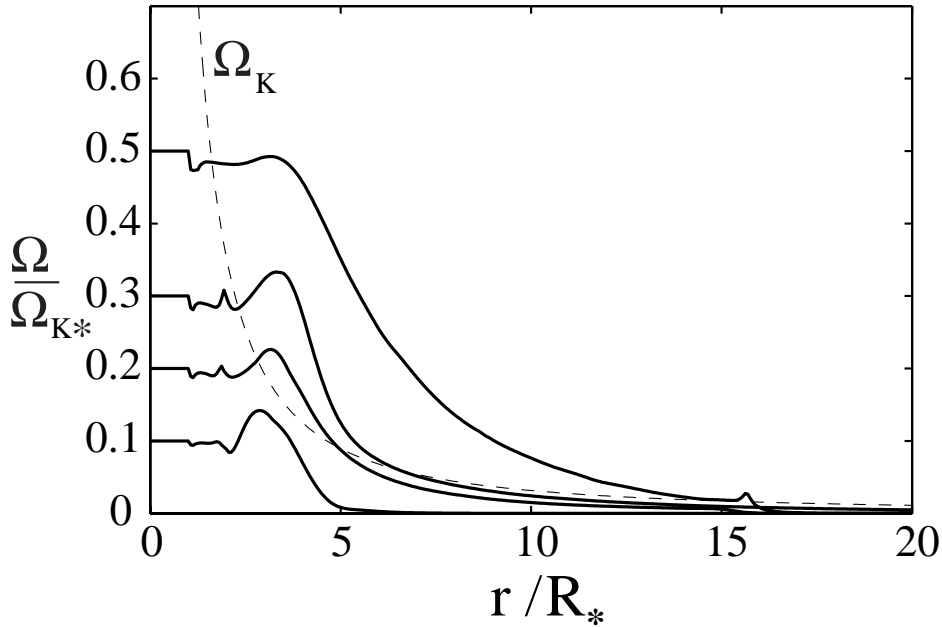} \caption{Angular velocity of matter
$\Omega=v_\phi/r$ versus $r$ in the equatorial region for
different angular velocities of the star. The dashed line
represents the Keplerian angular velocity $\Omega_K
=\sqrt{GM/r^3}$.}
\end{figure}

Figure 5a shows that the accretion rate to the star decreases as
the angular velocity of the star increases, $\dot{M}/\dot{M}_B
\propto \omega_*^{-1.0}$.
    Figure 5b shows that the accretion rate to the star decreases as
the star's magnetic moment increases, $\dot{M}/\dot{M}_B \propto
\mu^{-2.1}$.
     We have also done a number of simulation runs for different
magnetic diffusivities in the range $\tilde{\eta}_m =
10^{-6}-10^{-4.5}$, and from this we conclude that
$\dot{M}/\dot{M}_B \propto ({\eta}_m)^{0.7}$.

\section{Conclusion}

Axisymmetric magnetohydrodynamic simulations of Bondi accretion to
a rotating magnetized star in the propeller regime of accretion
have shown that:

 (1) A new regime of matter flow forms around a rotating star.
Matter falls down along the axis, but only a small fraction of the
incoming matter accretes to the surface of the star. Most of the
matter is expelled radially in the equatorial plane by the rotating
magnetosphere of the star. A low-density torus forms in the
equatorial region which rotates with velocity significantly larger
than the radial velocity. Large scale vortices form above and below
the equatorial plane.

 (2) The accretion rate to the star is much less than the Bondi accretion rate and
decreases as the star rotation rate increases ($\propto
\Omega_*^{-1.0}$), (b) as the star's magnetic moment increases ($
\propto \mu^{-2.1}$), and as the magnetic diffusivity decreases
[$\propto (\eta_m)^{0.7}$].

 (3) Because the accretion rate to the star is less than the Bondi rate, a shock
wave forms in our simulations and propagates outward. It has the
shape of an ellipsoid flattened along the rotation axis of the
star.

\section{Acknowledgements}

This work was supported by NSF grant AST 0307817 and by RFBR grant
05-02-17697. We thank Dr. V.V. Savelyev and Dr. Yu.M. Toropin for
the development of the original version of the MHD code used in
this work.

\begin{figure}
\plotone{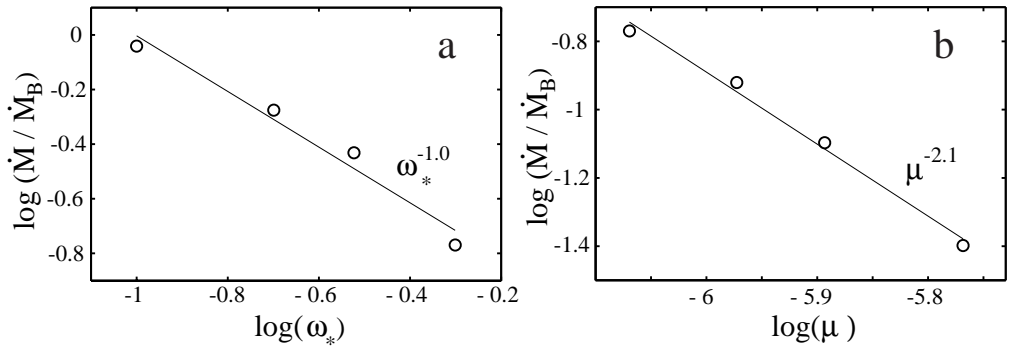} \caption{The left-hand panel ($a$) shows
the fraction of Bondi accretion rate reaching the star as a
function of the star's angular velocity
$\omega_*=\Omega_*/\Omega_{K*}$. The right-hand panel ($b$) shows
the dependence on the magnetic moment $\mu$ for $\Omega_*=0.5
\Omega_K$.}
\end{figure}

\begin {references}
\reference Bondi, H. 1952, MNRAS, 112, 195

\reference Davidson, K., \& Ostriker, J.P. 1973, ApJ, 179, 585

\reference Goldreich, P., \& Julian, W.H. 1969, ApJ, 157, 869

\reference Illarionov, A.F., \& Sunyaev, R.A. 1975. A\&A, 39, 185

\reference Lipunov, V.M. 1992, {\it Astrophysics of Neutron
Stars}, (Berlin: Springer Verlag)

\reference Romanova, M.M., Toropina, O.D., Toropin, Yu.M., \&
Lovelace, R.V.E. 2003, ApJ, 588: 400-407

\reference Shapiro, S.L., \& Teukolsky, S.A. 1983, ``Black Holes,
White Dwarfs, and Neutron Stars", (Wiley-Interscience)

\reference Shvartsman, V.F. 1970, Radiofizika, 13, 1852

\reference Toropin, Yu.M., Toropina, O.D., Savelyev, V.V.,
Romanova, M.M., Chechetkin, V.M., \& Lovelace, R.V.E. 1999, ApJ,
517, 906

\reference Toropina, O.D., Romanova, M.M., Toropin, Yu.M., \&
Lovelace, R.V.E. 2003, ApJ, 593: 472-480

\end {references}

\end{document}